\documentclass[12pt]{article}
\usepackage{epsfig}

\parskip=\medskipamount % space between paragraphs (TeX)
plus.3em minus.5em \abovedisplayshortskip=.5em plus.2em minus.4em
\renewcommand{\thesection}{\arabic{section}}

\catcode`@=11

\@addtoreset{equation}{section} \@addtoreset{equation}{subsection}
\def\theequation{\ifnum\value{section}=0 \arabic{equation}\ignorespaces
\else \ifnum\value{section}=-1 A.\arabic{equation}\ignorespaces
\else \ifnum\value{subsection}=0
\thesection.\arabic{equation}\ignorespaces \else
\thesection.\arabic{subsection}.\arabic{equation}\ignorespaces
                             \fi
                        \fi
                   \fi}

{\catcode`\'=\active \def'{{}^\bgroup\prim@s}}

\catcode`@=12

\newcommand{\bq}{\begin{equation}}
\newcommand{\be}{\begin{equation}}
\newcommand{\fq}{\end{equation}}
\newcommand{\ee}{\end{equation}}
\newcommand{\bqr}{\begin{eqnarray}}
\newcommand{\beqs}{\begin{eqnarray}}
\newcommand{\fqr}{\end{eqnarray}}
\newcommand{\eeqs}{\end{eqnarray}}

\newcommand{\rf}[1]{(\ref{#1})}

\def\bop#1{\setbox0=\hbox{$#1M$}\mkern1.5mu
    \vbox{\hrule height0pt depth.04\ht0
    \hbox{\vrule width.04\ht0 height.9\ht0 \kern.9\ht0
    \vrule width.04\ht0}\hrule height.04\ht0}\mkern1.5mu}
\begin{document}
\thispagestyle{empty}

\begin{flushright}
\begin{tabular}{l}
% TEP- \\
hep-th/0507171 \\
\end{tabular}
\end{flushright}

\vskip .6in
\begin{center}

{\bf A Format for Instantons and Their Characterizations}

\vskip .6in

{\bf Gordon Chalmers}
\\[5mm]
% {\em address \\
%      address \\
% Los Angeles, CA } \\

{e-mail: gordon@quartz.shango.com}

\vskip .5in minus .2in

{\bf Abstract}

\end{center}

A characterization of instanton contributions to gauge field theory is 
given.  The dynamics of instantons with the gluon field is given in terms of 
'classical' instanton contributions, i.e. on the same footing as tree 
amplitudes in field theory.  This paramaterization depends on the kinematics 
of the local instanton operators and their coupling dependence.  A 
primary advantage of this 'classical' formulation is that the weak-strong 
duality and its relations with the perturbative sector are possibly made more manifest.  

\vfill\break

\section{Introduction} 

The classical dynamics of gauge theory has been investigated for many years.  
Obstacles associated with the computation of these tree amplitudes have been 
circumvented for a variety of reasons, including spinor helicity, color 
ordering, analyticity, KLT relations with gravity, the string-inspired 
derivation, and the recent twistor formulation of classical gauge theory.  
The derivative expansion is useful for finding results in the quantum 
regime \cite{Chalmers1}-\cite{Chalmers16}.  

Potential simplifications based on a well-ordered formulation of the 
instanton contributions to an $n$-point scattering amplitude can be 
useful in finding the total correction to the quantum amplitudes.  These 
terms to the amplitude take on the form at $n$-point, 

\bqr 
f_{m,n}= g_{m,n}(g,N) \prod {1\over t_{i}^{[p],a_{i,p}}} e^{-m{4\pi\over g^2}+ 
 m{\theta\over 2\pi}} \quad K(\varepsilon_i,k_i) \ ,   
\label{instantonterms}
\fqr 
with $a_{i,p}$ labeling the power of the invariant (both negative and 
positive). 
The helicity content containing the factors $\varepsilon_i\cdot k_j$ 
and $\varepsilon_i\cdot \varepsilon_j$ is represented by the function 
$K(\varepsilon_i,k_j)$.  The kinematic invariants $t_{i}^{[p]}$ are 
defined by 

\bqr 
t_{i}^{[p]} = (k_i+k_{i+1}+\ldots + k_{i+p-1})^2 \ , 
\fqr 
for an ordering of the external legs in the cyclic fashion $(1,2,\ldots,n)$.  

The coefficients $g_{m,n}$ enter as a result of the integration of the 
non-trivial gauge field configuration at the instanton number $m$ and at 
$n$-point.  
The integration produces the kinematic invariants.  There is expected 
to be symmetry between the numbers $g_{m,n}$ at the various orders spanned 
$m$ and $n$. 

The classical scattering of the gauge field has much symmetry, which is 
not evident in the usual perturbative approach.  This symmetry is partly 
based on the collection $\phi_n$ of numbers as given in 
\cite{Chalmers4}-\cite{Chalmers5}.  
The $n$-point scattering can be formulated set-theoretically in terms of 
$n-2$ numbers which range from $2$ to $n$, for a sub-amplitude with the 
color-ordering from $1$ to $n$.  These numbers are such that the maximum 
number $n$ may occur $n-2$ times, with the minimum number $1$ not occuring 
at all.  The multiplicative times the numbers appear is denoted as $p_i$ 
(for $i=1,\ldots,n$) and the collection $p_1,\ldots,p_n$ completely label 
the amplitude.  (In another set theoretic form these numbers give way to a 
Hopf algebra of the tree diagrams \cite{ConnesKreimer}.)  All combinations 
of these numbers $\phi_n$ or $p_i$ generate the tree contributions to the 
gauge scattering, in a given color ordering of the sub-amplitudes.  

The suggestive form of these simple groups of numbers suggests that the 
'classical' (or semi-classical) contributions of the multi-instanton 
configurations to the scattering should also be labeled by a partition 
of the same numbers.  These partitions label the kinematic prefactors 
in \rf{instantonterms}.  
The symmetry of the classical form and that of the partitions labeling 
the instanton terms should be relevant to duality information in the 
gauge theory.  

In previous work the classical perturbative tree diagrams are used to 
find the full quantum scattering by sewing them together; the integals 
are performed and the primary complication is the multiplicity of the 
indices.  These tree diagrams are used as nodes, as essentially vertices.  
In this work the instanton contributions are formulated to achieve the 
same result; although the exact functional dependence is not yet known, 
the nodes and sewing may be formulated in the same manner as the 
perturbative results. The full quantum scattering, containing also 
the non-perturbative physics may be illustrated as in Figure 1.  The 
coefficients of the invididual vertices (labeled by $G_1$ and $G_2$) 

\bqr
V(G_1,G_2)=\sum_{n=0}^\infty \chi_n(G_1,G_2) z^n \qquad 
  z=e^{-{4\pi\over g^2}-i{\theta\over 2\pi}}
\fqr
are required to be determined, i.e. $\chi_n(G_1,G_2)$.  These numbers in 
principle are found from a unifying function via the pertinent number 
partitions that are similarly analyzed in the perturbative gauge context. 
The perturbative nodes required to find the scattering are given by 

\bqr 
V(G_p)= \chi(G_p) g^{m-2} \ , 
\fqr  
for an $m$-point tree represented by $\chi_n(G_p)$ (one of many trees 
required to specify an $m$-point classical amplitude). 

\begin{figure}
\begin{center}
\epsfxsize=12cm
\epsfysize=8cm
\epsfbox{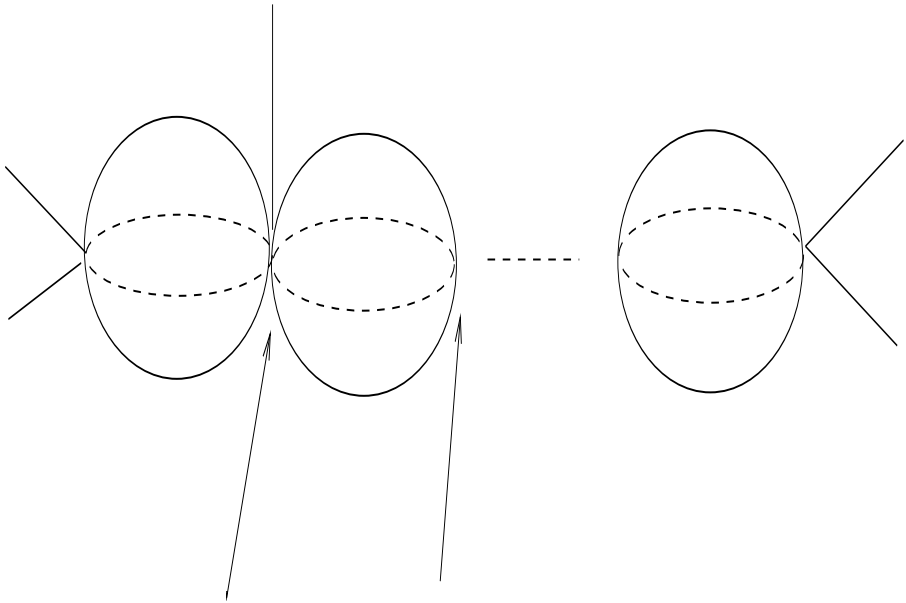}
\end{center}
\caption{ The product form solution to the recursive formulae defining 
the loop expansion.  The nodes are classical scattering vertices.  These 
vertices are both perturbative and non-perturbative}  
\end{figure}

The classical scattering is briefly reviewed in section 2.  A generalizaton 
based on the partitions of numbers which gives more kinematic expressions 
other than perturbative tree diagrams is presented in section 3.  The 
instanton analog is given in section 4.  A simple graphical interpretation 
of these results is presented in section 5. 

\vskip .2in 
\section{Classical Perturbative Scattering} 

Classical gauge theory scattering has been formulated in terms of the 
sets $\phi_n$ in \cite{Chalmers4}.  This representation of the gauge 
amplitudes has not included the spinor helicity technique, which with 
some further algebraic manipulations leads to maximally compact 
representations; some of the simplifications are due to the fact that 
the amplitudes have a 
symmetrical representation in terms of the sets of numbers $\phi_n$.  
The construction of the classical amplitudes in terms of these numbers 
is briefly reviewed.   

\vskip .2in
\noindent{\it Brief Review of the $\phi^3$ Labeling} 

The general scalar $\phi^3$ momentum routing of the propagators is 
presented.  The gauge theory tree amplitudes are also formulated with 
the scalar momentum routing; their tensor algebra can be found with the 
use of the string inspired formulation combined with the number partitions 
labeling the scalar graphs.  

A general $\phi^3$ scalar field theory diagram at tree-level is parameterized 
by the set of propagators and the momenta labeling them.  In a color 
ordered form, consider the ordering of the legs as in $(1,2,\ldots,n)$.  
The graphs are labeled by 

\bqr 
D_\sigma = g^{n-2} \prod {1\over t_{\sigma(i,p)}} \ , 
\label{phi3diagrams} 
\fqr 
and the Lorentz invariants $t_{\sigma(i,p)}$ are defined by $t_i^{[p]}$, 
 
\bqr  
t_i^{[p]} = (k_i+\ldots + k_{i+p-1})^2 \ .  
\label{momentainv}
\fqr 
Factors of $i$ in the propagator and vertices are placed into the prefactor of 
the amplitude.
The sets of permutations $\sigma$ are what are required in order to specify 
the individual diagrams.  The full sets of $\sigma(i,p)$ form all of the 
diagrams, at any $n$-point order.  These $\phi^3$ diagrams and their 
labeling, via the sets $\sigma$, are required in order to find the gauge 
theory amplitudes.   

The compact presentation of $\sigma$ can be obtained by a set of numbers 
$\phi_n$, discussed next, that generate the individual diagrams.  These 
sets are quite simple, and indirectly generate whats known as the Hopf 
algebra of the diagrams.  

First, the vertices of the ordered $\phi^3$ diagram are labeled so that 
the outer numbers from a two-particle tree are carried into the tree 
diagram in a manner so that $j>i$ is always chosen from these two numbers.  
The numbers are carried in from the $n$ most external lines. 

The labeling of the vertices is such that in a current or on-shell diagram the 
unordered set of numbers are sufficient to reconstruct the current; the set 
of numbers on the vertices are 
collected in a set $\phi_m(j)$.  For an $m$-point current there are 
$m-1$ vertices and hence $m-1$ numbers contained in $\phi_m(j)$.  These $m-1$ 
numbers are such that the greatest number may occur $m-1$ times, and must 
occur at least once, the next largest number may occur at most $m-2$ times 
(and may or 
may not appear in the set, as well as the subsequent ones), and so on.  The 
smallest number can not occur in the set contained in $\phi_m(j)$.  
Amplitudes are treated in the same manner as currents.  Examples and a 
more thorough analysis is presented in \cite{Chalmers4},\cite{Chalmers5}. 

Two example permutation sets pertaining to $4$- and $5$-point currents are 
given in \cite{Chalmers5}.  The five point amplitudes have sets of numbers 
such as $(5,5,5,5)$ and $(5,4,3,2)$, as an example.  

The numbers $\phi_n(j)$ are used to find the propagators in the 
labeled diagram.  The external momenta are $k_i$, and 
the invariants are found with the algorithm,     

\vskip .2in 
1) $i=\phi(m-1)$, $p=2$, then $l_{a_{m-1}}+l_{a_m}\rightarrow l_{m-1}$ 

2) $i=\phi(m-2)$, $p=2$, then $l_{a_{m-2}}+l_{a_{m-1}}\rightarrow l_{m-2}$ 

... 

$m-1$) $i=1$, $p=m$
\bqr \label{sigmarules} 
\fqr 
\vskip .2in 

\noindent The labeling of the kinematics, i.e. $t_i^{[p]}$, is direct from the 
definition of the vertices.  

Alternatively, the 
numbers $\phi_n(i)$ can be arranged into the numbers $(p_i,[p_i])$, in 
which $p_i$ is the repetition of the value of $[p_i]$; as an example, if the 
number $p_i$ equals zero, then $[p_i]$ is not present in $\phi_n$.  The  
numbers 
can be used to obtain the $t_i^{[q]}$ invariants without intermediate steps 
with the momenta.  The branch rules to determine the presence of  
$t_i^{[q]}$ is,  

\vskip .2in
0) $l_{\rm initial}=[p_m]-1$

1) 

$r=1$ to $r=p_m$  

${\rm if~} r + \sum_{j=l}^{m-1} p_j = [p_m]-l_{\rm initial}   
\quad {\rm then}~i= [p_m]   \quad q= [p_m] - l_{\rm initial}+1$ 

beginning conditions has no sum in $p_j$

2) 

${\rm else~} \quad l_{\rm initial}\rightarrow l_{\rm initial}-1$ : 
decrement the line number

$l_{\rm initial}>[p_{l}]$ else $l\rightarrow l-1$ : decrement the $p$ sum 

3) ${\rm goto}~ 1)$ 
\bqr  
\label{branchrules}
\fqr  
The branch rule has to be iterated to obtain all of the poles. 
The procedure uses the $\phi^3$ vertices and matches with the momentum 
flow to determine if a tree is present in a clockwise manner.  If not, 
then the external line number 
$l_{initial}$ is changed to $l_{initial}$ and the tree is checked again.  
The $i$ and $q$ are labels to $t_i^{[q]}$.  

The previous recipe pertains to currents and also on-shell amplitudes.  There 
are $m-1$ poles in an $m$-point current $J_\mu$ or $m-3$ in an $m$-point 
amplitude.  The comparison between amplitudes and currents is as follows: the 
three-point vertex is attached to the current (in $\phi^3$ theory), and then 
the counting is clear when the attached vertex has two external lines with 
numbers less than the smallest external line number of the current 
(permutations to other sets of $\phi_n$ does not change the formalism).  
There are $n-3$ poles are accounted for in the amplitude with $\phi_n$ and 
the branch rules.  

\vskip .2in 
\noindent{\it Brief Review of the Gauge Theory Labeling} 

The gauge theory contributions follow a similar labeling \cite{Chalmers4}, 
but with the 
added complexity of the kinematics in the numerator, such as $\varepsilon_i 
\cdot k_i$ and $\varepsilon_i\cdot \varepsilon_j$.  These pairings are 
determined set-theoretically from the integers in $\phi_n$, where the 
latter labels the momentum flow of the individual tree diagrams.

The $\kappa(a;1)$ and $\kappa(b;2)$ set of primary numbers used on 
can be found via the set of string inspired rules for the numerator 
factors, and define their individual contributions by, 

\bqr 
(-{1\over 2})^{a_1} ({1\over 2})^{n-a_2} 
\prod_{i=1}^{a_1} \varepsilon(\kappa(i;1))\cdot \varepsilon(\kappa(i;1))  
\fqr 
\bqr 
 \times \prod_{j=a_1+1}^{a_2} \varepsilon(\kappa(j;1)) \cdot k_{\kappa(j;2)} 
\times \prod_{p=a_2+1}^n k_{\kappa(p;1)} \cdot k_{\kappa(p;2)}  \ ,
\label{kappaterms}
\fqr 
together with the permutations of $1,\ldots,n$.  The permutations extract 
all possible combinations as explained in \cite{Chalmers4}.  

The form of the amplitudes is then expressed as $T_\sigma$ multiplying 
the propagators in \rf{phi3diagrams}.  $T_\sigma$ is given in \rf{kappaterms}, 
and the sets $\kappa$ are determined in \cite{Chalmers4} (using the string 
ispired rules for amplitudes).  The normalization is $i(-1)^n$.  The numbers 
$a_1$ and $a_2$ are summed so 
that $a_1$ ranges from $1$ to $n/2$, with the boundary condition 
$a_2\geq a_1+1$.  Tree amplitudes in gauge theory must possess at least 
one $\varepsilon_i\cdot\varepsilon_j$.  Also, the color structure is 
${\rm Tr} \left(T_{a_1}\dots T_{a_n}\right)$,  and the complete amplitude 
involves summing the permutations of $1,\ldots, n$.  

\vskip .2in 
\section{Product Form of Invariants}

The number partitions have been used to generate scalar $\phi^3$ graphs 
in \cite{Chalmers5}.  Their use is involved in generating the tensor 
algebra in gauge theory classical scattering as in \cite{Chalmers2} and 
\cite{Chalmers4}.  The general instanton vertex requires generally non-tree 
like combinations of the kinematic invariants, and so a generalization 
of these number partitions is required to generate their form.  The 
general combination of kinematics can also be used to label the quantum 
vertices in the scattering as they are not tree-like kinematically.  

In the previous section it was shown how certain partitions of numbers 
generate the non-vanishing coefficients $C_{i,p}$ in the scalar scattering.  
In perturbative gauge theory these coefficients are determined classically 
from the sets $\phi_n$ as in \cite{Chalmers1}-\cite{Chalmers5}.  

In a similar vein to the the perturbative gauge theory work, these 
numbers could be represented pictorially as in Figure 2.  Each 
set of numbers $\sigma(n)$ corresponds to deleted entries in the 
color ordered set $(1,2,\ldots,n)$ 

\bqr 
(1,2,\ldots, n) \quad (1,2;3,\ldots,n-2;n-1, n) \quad (1;2,3;4,\ldots,n) \quad 
\label{example}
\fqr 
The deleted entries are grouped together with other numbers in the set as 
the semi-colon indicates, but with the deleted entries neigboring the 
largest non-deleted number, such as $n-2$ with the 'deleted entries $3$ to 
$n-3$ in \rf{example}.  These sets of numbers are in analog to labeling 
all diagrams with legs grouped together as in the pair in Figure 2; the 
second diagram has legs 
four and five grouped (corresponding to a set $(1,2,3;4,5)$.  
In this manner the 
sets of numbers $C_{i,p;\sigma(n)}$ span all $\phi^3$ diagrams with ordered 
sets of legs.  The numbers $\sigma(n)$ can be put into one-to-one 
correspondence with the integers from $1$ to 

\bqr 
\sum_{i=1}^{n-1} {n!\over i!(n-i)!} = 2^{n-1}  \ ,   
\fqr 
as the number is either absent or present ($2^{n-1}$).  

The sets of numbers span all possible ordered pole structures in 

\bqr 
\prod C_{i,p;\sigma(n)} {1\over t_{i,p}}  \ .  
\label{instantonpoles}
\fqr 
As a trivial example, the second graph in Figure 2 could be $n$-point graph 
with the non-three last legs grouped with the leg $n$; the set $\sigma$ 
is $(4,\ldots,n-1)$ and the single invariant $s_{12}$ is obtained in the 
denominator.  In general it is possible to construct any combinations 
by grouping into two blocks these numbers, but in doing so the symmetry 
of the tree and non-tree diagrams is lost.  As in constructing the 
tree diagrams of both scalar and gauge field theory, the sets $\phi_n$ 
are very well-ordered sets of numbers and presumably this holds in the 
classical vertices of the instanton case as well.  

The $2^{m-1}$ possible leg orderings and topologies labeled by the 
individual $\phi_m$ sets produce all possible sets of pole 
terms; these sets $\phi_m$ at $m$-point together with the $2^{m-1}$ 
possibilities inherit a pseudo-Hopf structure.  The construction however 
is quite explicit.  
   
The coefficients $C_{i,p;\sigma(n)}$ span the graphs generating the 
pole terms in \rf{instantonpoles}.  A sample set of pole terms is 
given in Figure 2.  The perturbative sets of numbers $C_{i,p}$ 
is given by the sets $\phi_n$ as described in \cite{Chalmers5}.

\begin{figure}
\begin{center}
\epsfxsize=10cm
\epsfysize=7cm
\epsfbox{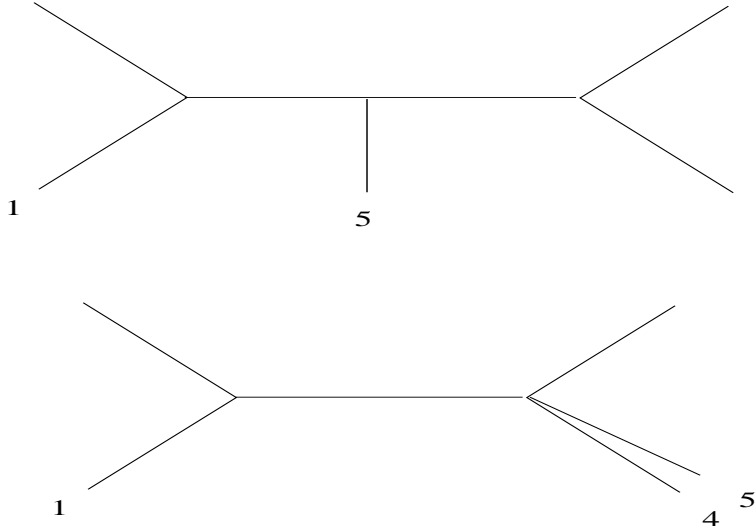}
\end{center}
\caption{This number partition, and the additional diagram, reproduces the 
function $1/s_{12}^2 s_{34}$.}  
\end{figure}

The labeling of the numbers $C_{i,p;\sigma(n)}$ is required to obtain 
the functions $g_{C_{i,p;\sigma(n)}}$ that multiply the individual terms.  
Assumed is that there is a group theory interpretation of the individual 
$g$ functions for a given set of terms at $n$-point order, spanned by 
the numbers $C_{i,p;\sigma(n)}$.  This group theory relevance should 
have an interpertation in terms of the classical perturbative scattering, 
together with the $2^{n-1}$ global possible interpretations.  

The determination of the vertix kinematics of both the perturbative 
and non-perturbative terms (i.e. instantons) in terms of the possible 
sets $\phi_n$ and $\phi_n,\sigma(n)$ might allow for a better understanding 
of the quantum duality from the classical configurations.  The full quantum 
scattering is obtained from both of these contributions, and any duality 
must manifest itself at the level of the tree-like nodes.  A kinematic 
duality in the partitions of the numbers, such as required from $\phi_n$ 
to generate the classical perturbative graphs, seems possible with the 
instanton contributions.  The kinematic structure of both are required 
to find for example, S-duality in the ${\cal N}=4$ supersymmetric gauge 
theory.  

\vskip .2in 
\section{Instanton Analogue} 

The perturbative scattering has been modeled through the use of classical 
tree diagrams sewn together to find the loop scattering.  
The non-perturbative scattering incorporating the $e^{-4\pi/g^2}$ 
effects requires the introduction of further terms in the nodes of 
Figure 1.  These further terms are modeled with the generalized kinematics 
of the previous sections together with the full coupling dependence.  
The reduction of the instanton contributions to the nodes is an algebraic 
simplification, and the full quantum scattering requires the iteration 
of the interactions as in Figure 1.  (The integrals and much of the 
tensor results have been performed in the latter.)  

The instanton vertices are sufficient, without logarithmic modifications, 
to deduce the approprate quantum analog of instantons interacting with the 
gluon field to any loop order.  For example, the instanton vertex at 
$n$-point can be defined to compensate the quantum recursion between 
the vertices and the result that is truely non-perturbative gauge theory; 
this compensation can be seen as adding irrelevant operators to fine tune 
a theory, but in this case, the classical instanton vertices are defined 
to agree with the quantum gauge theory instantons.  The 
appropriate matter is to deduce the kinematics and coupling dependence 
of the 'classical' instanton vertices.  

The instanton terms to an $n$-point amplitude recieve contributions 
from the $m$-instanton, and also at loop order $L$.  The most 
general term, with a coupling structure $f(g_{\rm YM})$ has the form, 

\bqr 
    f_{m,n;C}^{L}~ 
 e^{-m{4\pi\over g^2}+im{\theta\over 2\pi}} 
  \prod C_{i,p;\sigma(n)} {1\over t_i^{[p],a_{i,p}}} \ , 
\label{instantonform}
\fqr 
with the coupling dependence $f$ arising from a quantum corrected 
instanton moduli space; $a_{i,p}$ denotes general powers of the 
invariants, both positive and negative.  The helicity structure is,  

\bqr  
\prod \varepsilon_{\sigma(j)} \cdot k_{{\tilde\sigma}(j)} 
 \varepsilon_{\kappa(j)} \cdot \varepsilon_{{\tilde\kappa}(j)} \ . 
\fqr 
The coefficients $C_{i,p}$ are 
coefficients that take on the values $0$ and $1$.  They are set theoretic 
numbers that label the non-vanishing of the invariants $t_i^{[p]}$.  The 
logarithmic possibilities require another set of numbers $\tilde C$ to 
label, but the classical vertices do not require the soft dimensions so 
we neglect this notation (which is relevant in final results for quantum 
scattering).  The coefficients $C_{i,p}$ also generate the tensorial 
vectors $\sigma(j)$, $\tilde\sigma(j)$, and $\tilde\sigma'(j)$, which 
enter into the polarization inner products.

The function $f_{m,n;C}^L$ is expected to have an analogous 
determination as the coefficients describing perturbative gauge theory.  
(In the self-dual $N=4$ supersymmetric gauge theory context these functions 
are determined from the self-dual mapping $g\rightarrow 1/g$; one formulation 
involves the ansatz of the Eisenstein functions.) 

The pre-factor $f_{C_{i,p;a_{i,p}}}$, multiplying each term containing the 
products of $\varepsilon_i\cdot k_j$ and $\varepsilon_i\cdot\varepsilon_j$ 
has the coupling expansion at $m$-point,  

\bqr 
{\tilde f}^{(m)}_{C_{i,p;a_{i,p}}}= \sum b_n(C;g;m) 
  e^{-n{4\pi/g^2}+ni{\theta/2\pi}}   . 
\label{instexpansion}
\fqr 
The coefficients could in principle be determined from the appropriate 
expansion of a manifold. 

The vertices associated with every kinematic term in the series 
is used to find the full scattering in the gauge field, after 
including the perturbative terms.  The classical gauge field 
tree diagrams generates the classical vertices, as used in 
\cite{Chalmers1},\cite{Chalmers2} to generate the full quantum scattering.  
These tree diagrams are then added with the instanton vertices 
to find the full nodes.  The quantum theory is obtained by 
sewing both nodes together in the 'rainbow' graphs to determine 
the scattering, as illustrated in the Figure 1.  

\vskip .2in 
\section{Compact graphical form} 

The notation can be further simplified graphically with the use 
of two diagrams for the non-helicity specified amplitudes, and 
only one diagram for the helicity amplitude.  This graphical 
representation makes closer possible group theoretic and gometry 
in the interpretation of the instanton contributions, which could 
make a determination of the functions in \rf{instexpansion} easier.  

A labeled diagram $G_1$ with $m$ nodes ($m$-point vertex) is used 
to specify the contractions of the 
polarizations with the momenta, that is $\varepsilon_i\cdot k_j$ 
and $\varepsilon_i\cdot\varepsilon_j$.  The nodes label the ordered 
set of lines, such as $1,2,\ldots,n$.  There are two lines, with a 
dashed or solid line, that represent the contraction $\varepsilon_i\cdot 
\varepsilon_j$ or $\varepsilon_i\cdot k_j$, respectively.  The latter 
case requires an arrow orientation to label either the polarization or 
the momenta; the end of the arrow labels a momenta.  The figure is 
represented in figure 3.

\begin{figure}
\begin{center}
\epsfxsize=8cm
\epsfysize=6cm
\epsfbox{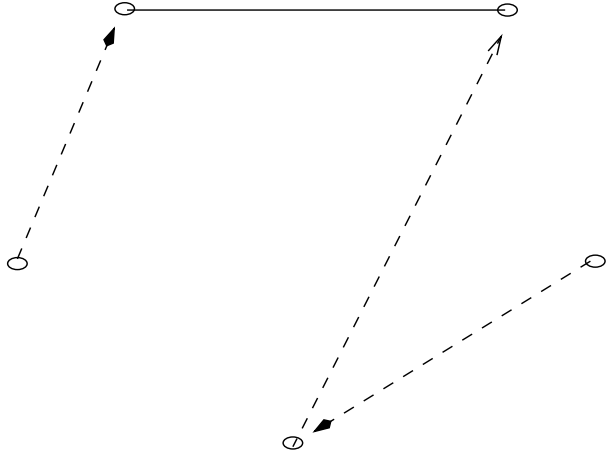}
\end{center}
\caption{The parameterization of the node kinematics.  The dashed and 
solid lines represent contractions of the polarizations with the 
momenta.  The circles are the nodes $1$ to $5$.}  
\end{figure}

\begin{figure}
\begin{center}
\epsfxsize=8cm
\epsfysize=6cm
\epsfbox{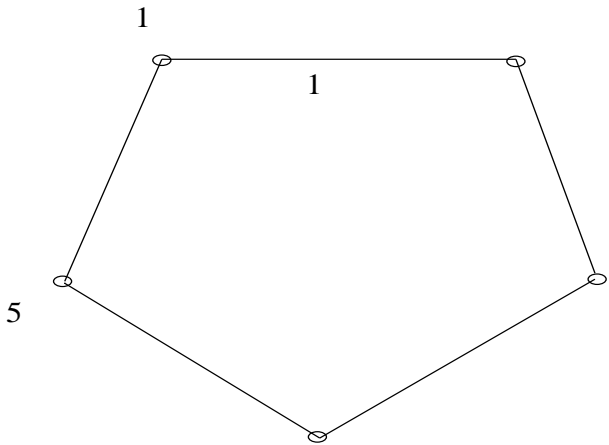}
\end{center}
\caption{The second diagram represents the contractions $(k_i+\ldots 
+ k_{i+j-1})^2$; numbers represent the powers of these invariants.  
This graph represents $\prod_{i=1}^5 1/s_{i,i+1}$.}  
\end{figure}

The second diagram $G_2$ is used to represent the various momentum 
invariants in the denominator and numerator.  The line from node 
$i$ to node $j$ labels the invariant $(k_i+\ldots +k_{i+j-1})^2$, 
and the index on the line represents its power; for example, the 
index of $1$ is a denominator and $-2$ is a double power in the 
numerator.  This diagram is represented in figure 5.  

Together these two diagrams label an individual contribution to 
the instanton vertex in figure 4.  The functional form of the 
coupling dependence are given by the equation, 

\bqr 
V^{(m)}(G_1,G_2)=\sum_{n=0}^\infty \chi_n^{(m)}(G_1,G_2;g) 
 e^{-n{4\pi\over g^2}-ni {\theta\over 2\pi}} \ .  
\fqr 
The graphical illustration alludes to a geometric derivation of the 
components $\chi_n(G_1,G_2)$.  Indeed, writing the 'classical' 
multi-instanton contributions as 

\bqr 
z=e^{-{4\pi\over g^2}- i{\theta\over 2\pi}} \ , 
\fqr 
generates a holomorphic function, 

\bqr
V^{(m)}(G_1,G_2)=\sum_{n=0}^\infty \chi_n^{(m)}(G_1,G_2;g) z^n \ ,  
\fqr 
similar to a K\"ahler potential.  Presumably specifying this function 
through its analytic properties, based on the symmetries of the graphs 
$G_1$ and $G_2$, generates the instanton contributions.  The latter 
allows the perturbative scattering of the quantum gauge theory to 
incorporate the non-perturbative terms.  

In a spinor helicity format the two diagrams can be reduced to only 
one diagram $G_s$.  The lines are arrows, with each one either dashed 
or solid, representing the inner products $\langle ij\rangle$ or 
$[ij]$.  The arrow specifies the orientation from $i$ to $j$ in the 
graph with $n$ nodes.

\begin{figure}
\begin{center}
\epsfxsize=8cm
\epsfysize=6cm
\epsfbox{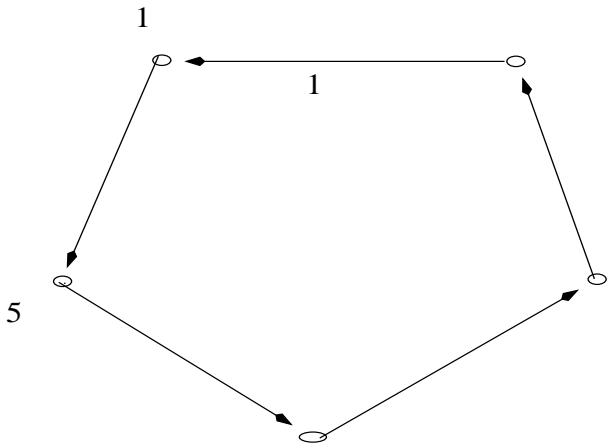}
\end{center}
\caption{The parameterization of the helicity basis nodes.  The 
numbers represent the powers of the invariants $\langle ij\rangle$ 
and $[ij]$. This graph represents $\prod_{i=1}^5 1/\langle i+1,i\rangle$}  
\end{figure}

\vskip .2in 
\section{Discussion} 

The classical scattering within gauge theory has a simple 
formulation in terms of sets of numbers $\phi_n$ that elucidates 
the instanton contribution formulation.  There are simplifications, 
further due to the spinor helicity implementation in this approach 
which are also algebraic.  It seems that a simple program 
will generate all contributions to obtain the full quantum amplitudes, 
with both perturbative and non-perturbative corrections.   

The sets of numbers $\phi_n$, or $p_i$, formulate a direct representation 
of the non-helicity format classical amplitudes.  The corresponding sets 
of numbers $\tilde\phi_n$, and $p_{i,n}$ generate the instanton kinematics 
at leading order.  The coefficients $V^{(n)}(G_1,G_2)$ are expected to follow 
a similar number theoretic, and an accordant differential representation.  
The group theory in association with these numbers should lead to a full 
determination of the instanton contributions to the $n$-point amplitude.   
  
The instanton interactions with the gauge bosons are modeled with the 
use of classical interaction terms entering into the sewing relations, 
the latter of which generate the full quantum theory amplitudes.  
Duality transformations at the basic level should operate only on these 
vertices in the coupling constant, as full transformations are repetitive 
on the individual nodes.  This property possibly simplifies the 
non-perturbative $g\rightarrow 1/g$ transformations within the quantum 
dynamics, 
when the appropriate transformations of the fields between the nodes are 
given such as a straight gluon to gluon map, or those in various theories.  
The appropriate transformations might generate the instantonic vertices 
and the non-perturbative contribution to the quantum theory.  

\vfill\break


\begin{thebibliography}{99}

\bibitem{Chalmers1} 
G. Chalmers, {\it Quantum Solution to Scalar Field Models}, physics/0505018.  

\bibitem{Chalmers2} 
G. Chalmers, {\it Quantum Gauge Amplitude Solutions}, physics/0505077.  

\bibitem{Chalmers3} 
G. Chalmers, in preparation. 

\bibitem{Chalmers4} 
G. Chalmers, {\it Tree Amplitudes in Gauge and Gravity Theories}, physics/0504219.  

\bibitem{Chalmers5}, 
G. Chalmers, {\it Tree Amplitudes in Scalar Field Theories}, physics/0504173.  

\bibitem{Chalmers6} 
G. Chalmers, {\it Inversions of Effective Action in Condensed Matter Models}, 
physics/0506012.  

\bibitem{Chalmers7}
G. Chalmers, {\it Derivation of Quantum Field Dynamics}, physics/0503062.

\bibitem{Chalmers8}
G. Chalmers, {\it Masses and Interactions of Nucleons Quantum Chromodynamics}, 
physics/0503110.  

\bibitem{Chalmers9}
G. Chalmers, {\it Comment on the Riemann Hypothesis}, physics/0503141.

\bibitem{Chalmers10}
G. Chalmers, {\it $N=4$ Supersymmetric Gauge Theory in the Derivative 
Expansion}, hep-th/0209088.

\bibitem{Chalmers11}
G. Chalmers, {\it Gauge Theories in the Derivative Expansion}, hep-th/0209086.

\bibitem{Chalmers12}
G. Chalmers, {\it Scalar Field Theory in the Derivative Expansion}, 
hep-th/0209075.

\bibitem{Chalmers13}
G. Chalmers, {\it M Theory and Automorphic Scattering}, Phys.\ Rev.\ D 
{\bf 64}:046014 (2001).

\bibitem{Chalmers14}
G. Chalmers, {\it On the Finiteness of $N=8$ Supergravity}, hep-th/0008162.

\bibitem{Chalmers15}
G. Chalmers and J. Erdmenger, {\it Dual Expansions of $N=4$ super Yang-Mills 
theory via IIB Superstring Theory}, Nucl.\ Phys.\ B {\bf 585}:517 (2000), 
hep-th/0005192.

\bibitem{Chalmers16}
G. Chalmers, {\it S and U-duality Constraints on IIB S-Matrices}, Nucl.\ 
Phys.\ B {\bf 580}:193 (2000), hep-th/0001190.

\bibitem{ConnesKreimer} 
A. Connes, D. Kreimer, {\it Renormalization in Quantum Field Theory and the 
Riemann-Hilbert Problem}, Comm.\ Math.\ Phys.\ {\bf 210}:249 (2000), 
hep-th/9912092; D. Kreimer, {\it Structures in Feynman Graphs: Hopf 
Algebras and Symmetries}, 
Proc.\ Symp.\ Pure\ Math.\ {\bf 73}:49 (2005), hep-th/022110 ; 
{\it Combinatorics of (Perturbative) Quantum Field Theory}, Phys.\ Rept.\ 
{\bf 363}:387 (2002), hep-th/0010059.   



\end{thebibliography}
\end{document}